\documentclass[linenumbers,graybox]{svmult}
\usepackage{type1cm}       
\usepackage{makeidx}         
\usepackage{graphicx}                                 
\usepackage{multicol}        
\usepackage[bottom]{footmisc}
\usepackage{newtxtext}        
\usepackage{newtxmath} 

\usepackage{hyperref}
\usepackage{multirow}
\usepackage{float}
\usepackage{lineno}
\usepackage{xspace}
\usepackage{float}

\usepackage{lineno, blindtext}

\makeindex                                    

\begin{document}

\title*{Insight into Multiple Partonic Interactions and Production of Charmonia in p+p Collisions at the LHC energies}
\author{Raghunath Sahoo{$^*$}, Dhananjaya Thakur, Sudipan De and Soumya Dansana$^\dagger$} 
\institute{Discipline of Physics, School of Basic Sciences, Indian Institute of Technology Indore, Indore- 453552, INDIA, {$^*$}Presenter \email{Raghunath.Sahoo@cern.ch}, $^\dagger$ Department of Physical Sciences, Indian Institute of Science Education and Research, Kolkata-741246, INDIA}


\maketitle

\abstract{At the LHC energies, the underlying observables are of major topic of interest in high multiplicity $p+p$ collisions. Multiple Parton Interactions (MPIs) is one of them, in which several interactions occur in a single $p+p$ collision. It is believed that MPI is the main reason behind the high multiplicity in $p+p$ collisions at the LHC. It was believed that MPI has only effect to the soft particle production, but recent ALICE result reveals that it can also affect the hard-particle production. In such case, the self normalized yield of heavy particle like $\rm J/\psi$ show increase trend with event multiplicity. In the present contribution, we discuss the energy and multiplicity dependence of charmonium production to understand the effects of MPI on charmonium production. }

\section{Introduction}
\label{intro}
\begin{figure}[H]
\begin{center}
\includegraphics[width=15pc]{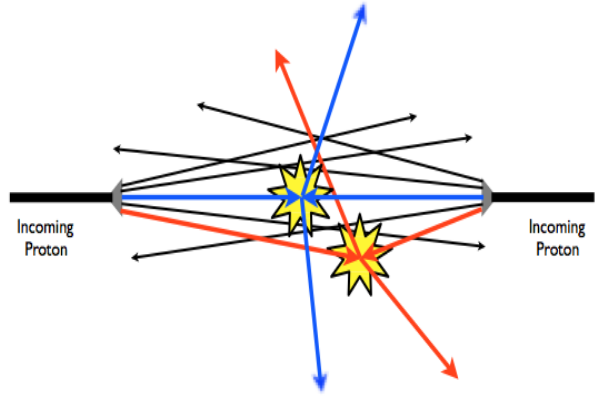}
\includegraphics[width=10pc]{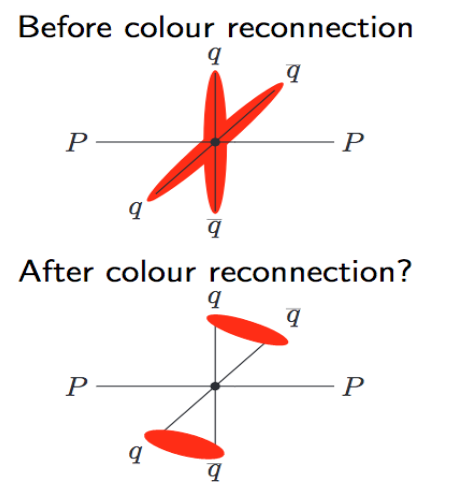}
\caption{Left: Event structure of a simple $p+p$ collision showing MPI. Right: An illustrative example of color reconnection mechanism with string minimisation.}
\label{fig:eventstructure}
\end{center}
\end{figure}

The understanding of event structure in hadronic collisions is a very challenging task. It is said that the physics associated with it is sub divided into a number of components, like hard central interactions, fragmentation of beam remnant, multi-partonic interactions (MPI), and initial and final state radiation (ISR and FSR) etc. Technically, this is called as underlying event, which is the sum of all the processes that build up the final hadronic state in a collision. Among all, MPI is of great interest at the LHC energies. Earlier, it was thought that MPI can only affect soft-particle production. But, recent result of production of heavy particle like D-meson and J/$\psi$ as a function of charged-particle multiplicity at  $\sqrt{s}$ = 7 TeV and 13 TeV~\cite{Abelev:2012rz, Weber:2017hhm, Adam:2015ota}  reveal that it has also effect on hard-particle production. ALICE experiment has observed a linear increase of open-charm and J/$\psi$ production as a function of multiplicity for $\sqrt{s}$ = 7 TeV.  Preliminary result of J/$\psi$ via di-electron channel at $\sqrt{s}$ = 13 TeV also shows a faster than liner increase of J/$\psi$ production with charged-particle multiplicity. QCD inspired models like PYTHIA6 could not explain the behavior as MPI process therein can only affect soft-processes.  An updated version of PYTHIA has been proposed, PYTHIA8~\cite{pythia8html}, where MPI plays an important role in the production of heavy quarks like charm and bottom. PYTHIA8 describes the increasing trend of heavy-flavor versus multiplicity at $\sqrt{s}$ = 7 and 13 TeV. Along with MPI, color reconnection (CR) is an important ingredient in PYTHIA8, which describes the interactions between color field during hadronisation. CR is expected to occurs in a significant rate at the LHC due to high number of color partons from MPI and parton shower. As we do not have J/$\psi$ versus multiplicity results for all the LHC energies, we have made an attempt to study energy and multiplicity dependence of J/$\psi$ production using PYTHIA8. In particular, we have studied the effect of MPI and CR on the production of J/$\psi$ and its behavior with respect to charged-particle multiplicity and $\sqrt{s}$.
\newline
\newline
\textbf{The Multiple Parton Interaction and J/$\psi$ production}
\newline
\newline
At the LHC with very high $\sqrt{s}$, the number of interactions in proton+proton collisions depend on the impact parameter (b), where proton is thought to be an extended object. Therefore matter distribution inside hadron (proton) is introduced. So small impact parameter leads to large multiplicity and hence more MPIs. At the LHC, a new QCD regime can be reached, where MPIs occur with high rates, due to unprecedentedly high parton densities in the colliding hadrons. Hence MPI, in particular can be defined as soft or hard interaction, which can happen in parallel in a single $p+p$ collision. The very first interaction is hard in particular and the subsequent collisions can be semi-hard or soft.  A schematic picture of MPI structure is shown in the Fig.~\ref{fig:eventstructure} (left) for a better visualization.

There is always a finite chance to produce heavy particle in first $2\rightarrow2$ collisions. But, The second hard can produce heavy quarks like charm and bottom, if it is hard enough and this leads to a correlation between heavy-flavor and charged-particle multiplicity. If former is true, then the self normalized yield of J/$\psi$ with respect to minimumbias as a function of charged-particle multiplicity should show a decreasing trend, which is observed in PYTHIA6 ~\cite{Abelev:2012rz}. This is because, PYTHIA6 consider $2\rightarrow2$ interaction in hadronic collisions. If later is true, then only we can observe an increasing trend of self normalized J/$\psi$ yield versus multiplicity.

\section{Analysis procedure}
\label{sec:expt}
Many of the models use to explain physics from $p+p$ collisions have MPI  processes in different ways. In our current study, we have used PYTHIA 8.2 tuned 4C~\cite{atlaspubnote}. As discussed in the introduction section, to achieve the objective of the study, we have included the varying impact parameter (MultipartonInteractions:bProfile=3) to allow all incoming partons to undergo hard and semi-hard interactions as well. We have used the MPI- based scheme of color reconnection (CR). In this scheme, the produced partons undergo a reconnection in which partons from lower-$p_{T}$ MPI systems are added to the dipoles defined by the higher-$p_{T}$ MPI system in such a way that minimizes total string length. An illustration to this is shown in a cartoon in Fig. 1 (right). After the color reconnection, all the produced partons, connected with strings, fragment into hadrons via the Lund string model.


We have performed this study by simulating the inelastic, non-diffractive component of the total cross section for all hard QCD processes (HardQCD: all = on), which includes the production of heavy quarks.  A $p_{T}$ cut of 0.5 GeV/c is used to avoid the divergences of QCD processes in the limit $p_{T}\rightarrow0$. The charged-particle multiplicity measurement has been performed in the mid-rapidity ($|y|<1.0$), where as J/$\psi$ has been measured at forward rapidity ($2.5 < y <4.0$), to cope with the ALICE measurement. Here to be noted that we are not directly studying yield of J/$\psi$ versus charged-particle multiplicity, rather self-normalized yields, where the charged-particle and J/$\psi$ yield in multiplicity bins are normalized to it's minimumbias yield. The purpose of studying this kind of ratio is to see, how the physics of multiplicity classes are different from that of minimum bias. Hence one can comment on MPI and other UE observables. The relative charged-particle multiplicity yield is defined as $\rm N_{ch}/<N_{ch}>$. Where $\rm N_{ch}$ is the mean of the charged-particle multiplicity in a particular bin and $<\rm N_{ch}>$ is the mean of the charged-particle multiplicity in minimum- bias events. The charged-particle multiplicity has been sliced taking care of significant number of J/$\psi$, and the self-normalized J/$\psi$ yield is calculated as,
\begin{equation}
\begin{aligned}
\label{eq:jpsi_relativeyield}
\frac{\frac{dN_{J/\psi}}{dy}}{<\frac{dN_{J/\psi}}{dy}>} = \frac{N_{J/\psi}^{i}}{N_{J/\psi}^{\rm integrated}} \times \frac{N_{\rm MB}^{\rm integrated}}{N_{\rm MB}^{i}}
\end{aligned}
\end{equation} 

where ($N_{J/\psi}^{i}$, $N_{\textrm{J}/\textrm{$\psi$}}^{ \textrm{\rm integrated}}$)  and   ($N_{\textrm{MB}}^{i}$, $N_{\textrm{MB}}^{\rm integrated}$) are the corrected number of J/$\psi$ and number of minimum bias events in $i^{th}$ multiplicity bin and integrated multiplicity bins, respectively. The detailed about the uncertainty calculation of each quantity can be found in Ref.~\cite{Thakur:2017kpv}. These uncertainties are propagated using the standard error propagation formula. We have tried to do multiplicity binning in such a way that, we can do direct comparison with experimental results of ALICE. 

\begin{figure}[H]
\begin{center}
\includegraphics[width=15pc]{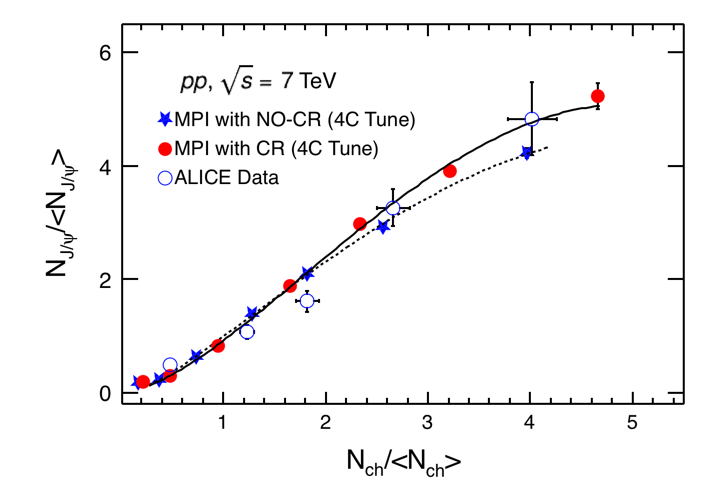}
\caption{Relative J/$\psi$ yield as a function of relative charged-particle multiplicity at  $\sqrt{s}$ = 7 TeV at the forward rapidity ($2.5 < y <4.0$). ALICE data \cite{Abelev:2012rz} and PYTHIA8 (with CR/ w/o CR) comparison is shown. The lines are the fitted curves using the percolation model inspired function \cite{Thakur:2017kpv}.}
\label{fig:7tev}
\end{center}
\end{figure}

To check the compatibility of the PYTHIA8 and experimental data, first we compare the basic distribution like integrated transverse momentum and rapidity spectra with experiment using the same phase space cuts. We found a good agreement between Monte Carlo (MC) and experimental data. Rapidity spectra gives around 1$\%$ maximum deviation. Where as for $p_{T}$-spectra the maximum deviation is $\sim$ 50-60$\%$ for certain $p_{T}$ bins, otherwise in most of the $p_{T}$ bins, the deviation is around 10-20$\%$ for $\sqrt{s}$ = 0.9, 2.76, 5.02, 7 and 13 TeV. After performing these compatibility studies, we compare the relative yield of J/$\psi$ as a function of charged particle multiplicity for available experimental data at  $\sqrt{s}$ = 7 TeV. It can be seen from the Fig.\ref{fig:7tev}, that experimental data are well described by the MC.

These measurements provide us the confidence to extend the study of quarkonia production using PYTHIA8 in $p+p$ collisions at LHC energies and perform the multiplicity and energy dependence analysis.

\section{Results}
\label{sec:result}
After performing all the feasibility tests as discussed in the previous section, we have extended the event multiplicity dependence of J/$\psi$ production to all the LHC energies $\sqrt{s}$ = 0.9, 2.76, 5.02, 7 and 13 TeV. 
\subsection{Multiplicity dependence study of J/$\psi$ production}
\label{subsec1}
\begin{figure}[H]
\begin{center}
\includegraphics[width=12pc]{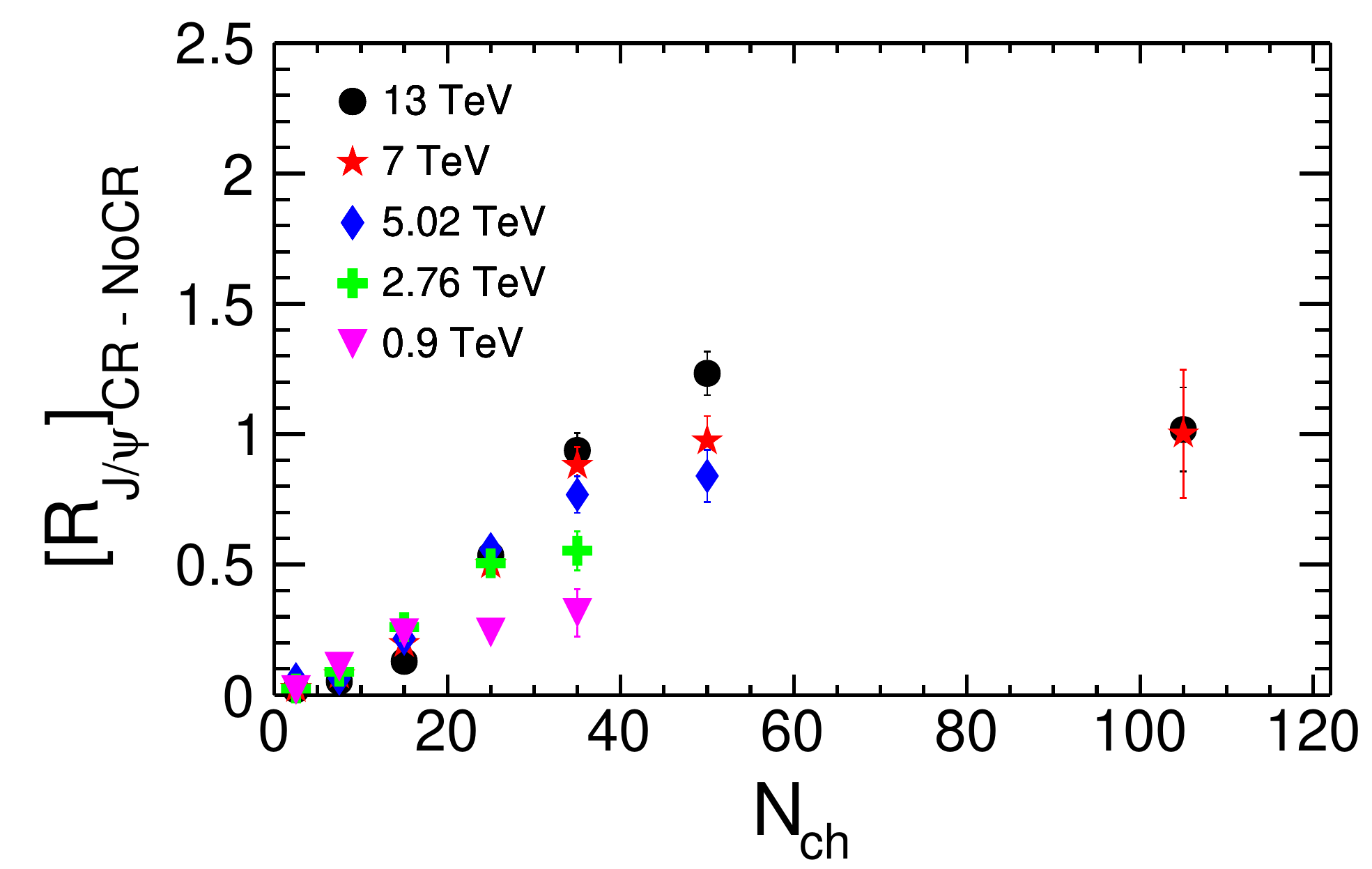}
\caption{The difference of relative J/$\psi$ yield, $\rm R_{J/\psi}$ with and without CR as a function of $\rm N_{ch}$ at the LHC energies.}
\label{fig:cr_nocr}
\end{center}
\end{figure}

We have simulated J/$\psi$ at all the LHC energies using 4C tuned PYTHIA8 and studied the energy dependence behavior to understand effect of MPI on J/$\psi$ production. This study is performed with two tunes of PYTHIA8: CR and No-CR, to see the final state effects on J/$\psi$ production. To understand the CR effects on J/$\psi$ production quantitatively, we have subtracted the yield of relative J/$\psi$ production with no-CR from the yield with CR and plotted it as a function of charged-particle multiplicity for different center-of-mass energies, which is shown in Fig.~\ref{fig:cr_nocr}.

It is found that the difference of the relative J/$\psi$ yield between with CR and without CR increases with charged-particle multiplicity as well as with increasing energy. At LHC energies due to the high density of colored partons, there is a substantial degree of overlap of many colored strings in the position and momentum phase space. Hence, there is a higher probability of color reconnection. The partons from two different MPIs can reconnect via color strings with the minimization of the string length as discussed in the previous section. This study reveals that with the increase of MPIs the probability of color reconnection increases and hence the probability to combine charm and anticharm quark becomes higher, thereby producing a higher number of J/$\psi$ particles.

\subsection{Energy dependence study of J/$\psi$ production} 
\label{subsec2}
\begin{figure} [H]
\begin{center}
\includegraphics[width=12pc]{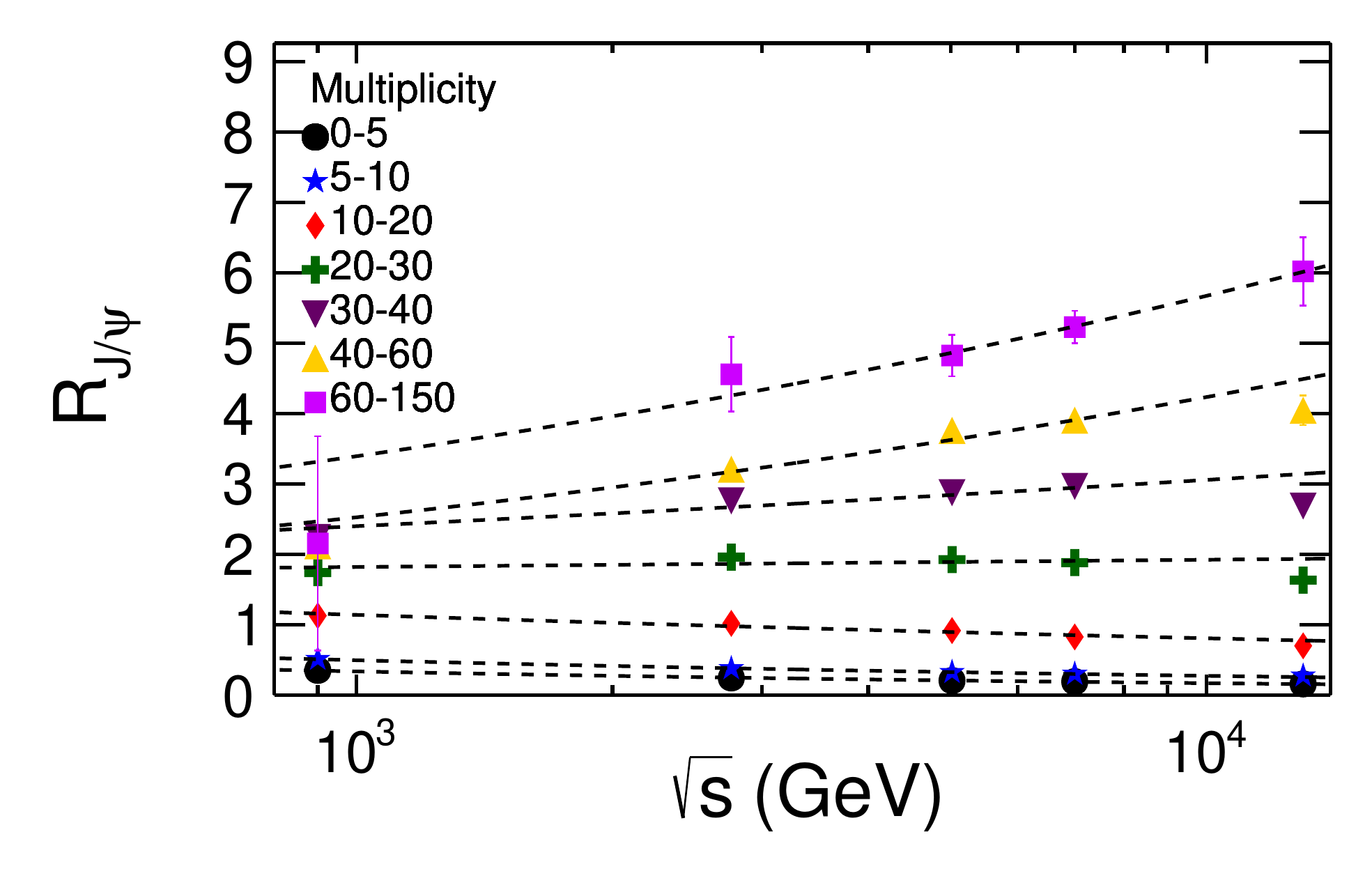}
\includegraphics[width=12pc]{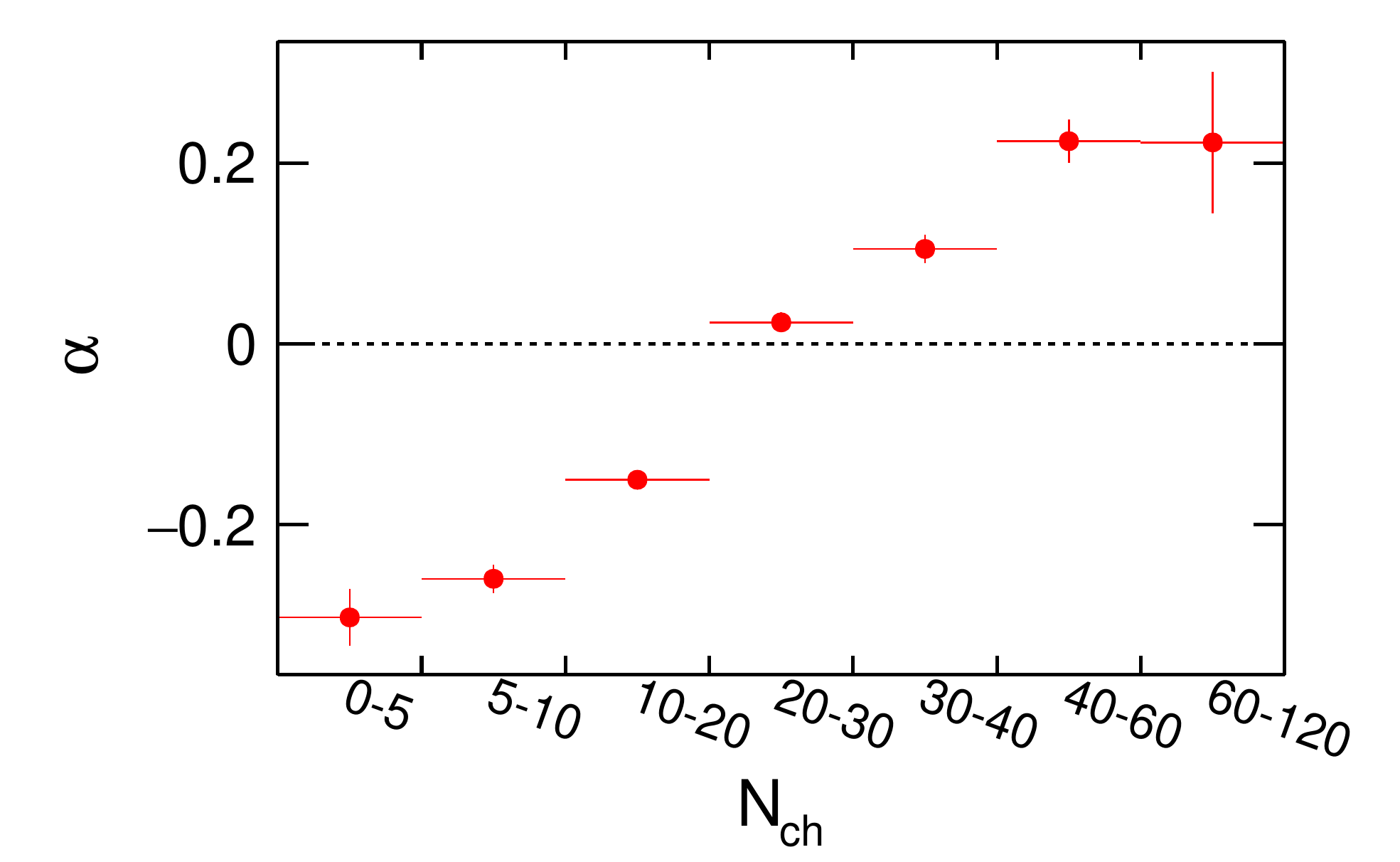}
\caption{Left panel: Relative J/$\psi$ yield as a function of $\sqrt{s}$, using PYTHIA8 with CR. The dashed lines are the phenomenological $y=Ax^{\alpha}$ fitting. Right panel: The fitting parameter ($\alpha$) as a function of charged-particle multiplicity.}
\label{fig:subsec2}
\end{center}
\end{figure}

Let's now explore the effect of MPI and CR on charmonia production at various collision energies. Figure~\ref{fig:subsec2} shows the relative J/$\psi$ yield as a function of the center-of-mass energy for different charged-particle multiplicity bins using PYTHIA8 with CR. It is found that the relative J/$\psi$ yield increases with $\sqrt{s}$. 

We have performed a  quantitative study by fitting the results with a phenomenological function, $y=Ax^{\alpha}$ , where A and $\alpha$ are the parameters. The parameter, $\alpha$, represents the rate of increase of relative J/$\psi$ as a function of center-of-mass energy for a particular multiplicity bin. From the left panel of Fig.~\ref{fig:subsec2}, the $\alpha$-parameter is found to increase with multiplicity. The values of $\alpha$ are negative up to $(10-20)$ multiplicity bins and become positive towards higher-multiplicity bins. This indicates that MPI effects dominate for J/$\psi$ production for $\rm N_{ch} > 20$. 
%
%
%

\section{Summary} 
\label{sec:sum}   
In this contribution, energy and multiplicity dependence of J/$\psi$ production has been presented using 4C tuned PYTHIA8. We have summarized below the important points drawn from the study.

 \begin{itemize}
 \item The difference between the relative J/$\psi$ yield with CR and without CR increases with charged-particle multiplicity as well as with center-of-mass energy. As the difference is very less, one can infer that final state effect has very less contribution to J/$\psi$ production.

 \item The relative J/$\psi$ yield as function of $\sqrt{s}$ shows a monotonic increase, where as for $\rm N_{ch}\le 20$, the behavior is opposite. This hints for dominance of MPI to the J/$\psi$ production from $\rm N_{ch} \ge 20$.
 
\end{itemize}

The present studies are important in view of the interesting properties shown by high-multiplicity events in $p+p$ collisions at the LHC energies. It will be very interesting to get the experimental measurements, which can help to explore more into the multiplicity dependence of quarkonia production. This work has appeared as a regular publication in Ref. \cite{Thakur:2017kpv}.

{}

\end{document}